\documentclass[11pt]{emulateapj}
\usepackage{epstopdf}

\usepackage{epsf}
\usepackage{natbib}
\usepackage{url}
\usepackage{indentfirst}
\usepackage{amsmath}
\usepackage{cases}
\usepackage{graphicx}

\def\jcap{Journal of Cosmology and Astroparticle Physics}
\def\pasa{Publications of the Astronomical Society of Australia}

\begin{document}

\title{Icecube non-detection of  GRBs: Constraints on the fireball properties }

\author{Hao-Ning He\altaffilmark{1,2,3}, Ruo-Yu Liu\altaffilmark{1,2}, Xiang-Yu Wang\altaffilmark{1,2},
Shigehiro Nagataki\altaffilmark{3}, Kohta Murase\altaffilmark{4},
Zi-Gao Dai\altaffilmark{1,2} } \altaffiltext{1}{School of
Astronomy and Space Science, Nanjing University, Nanjing, 210093, China;
haoninghe@nju.edu.cn, ryliu@nju.edu.cn, xywang@nju.edu.cn}
\altaffiltext{2}{Key laboratory of Modern Astronomy and Astrophysics
(Nanjing University), Ministry of Education, Nanjing 210093, China}
\altaffiltext{3}{Yukawa Institute for Theoretical Physics, Kyoto
University, Oiwakecho, Kitashirakawa, Sakyoku, Kyoto 606-8502,
Japan} \altaffiltext{4}{Department of Physics, Center for Cosmology
and AstroParticle Physics, The Ohio State University, Columbus, OH
43210, USA}

\begin{abstract}
The increasingly deep limit on the neutrino emission from gamma-ray
bursts (GRBs) with IceCube observations has reached the level that
could put useful constraints on the fireball properties. We first
present a revised analytic calculation of the neutrino flux, which
predicts a flux an order of magnitude lower than that obtained by
the IceCube collaboration. For benchmark model parameters (e.g. the
bulk Lorentz factor is $\Gamma=10^{2.5}$, the observed variability
time for long GRBs is $t_{\rm v}^{\rm ob}=0.01{\rm s}$ and the ratio
between the energy in accelerated protons and in radiation is
$\eta_p=10$ for every burst) in the standard internal shock
scenario, the predicted neutrino flux from 215 bursts during the
period of the 40-string and 59-string configurations is found to be
a factor of $\sim3$ below the IceCube sensitivity. However, if we
accept the recently found inherent relation between the bulk Lorentz
factor and burst energy, the expected neutrino flux increases
significantly and the spectral peak shifts to lower energy. In this
case, the non-detection then implies that the baryon loading ratio
should be $\eta_p\la10$ if the variability time of long GRBs is
fixed to $t_{\rm v}^{\rm ob}=0.01{\rm s}$. Instead, if we relax the
standard internal shock scenario but keep to assume $\eta_p=10$, the
non-detection constrains the dissipation radius to be $R\ga
4\times10^{12}{\rm cm}$ assuming the same dissipation radius for
every burst and benchmark parameters for fireballs. We also
calculate the diffuse neutrino flux from GRBs for different
luminosity functions existing in the literature. The expected flux {
exceeds} the current IceCube limit for some luminosity functions,
and thus the non-detection constrains $\eta_p\la10$  { in such
cases} when the variability time of long GRBs is fixed to $t_{\rm
v}^{\rm ob}=0.01{\rm s}$.
\end{abstract}


\section{Introduction}
GRBs have been proposed as one of the potential sources for
ultra-high energy cosmic rays (UHECRs) with energy up to
$>10^{20}{\rm eV}$ (e.g. Waxman 1995; Vietri 1995; Waxman $\&$ Bahcall 2000;
Dai $\&$ Lu 2001; Dermer 2002; Murase et al. 2006), given the hypothesis that the
fireball composition is proton-dominated and the protons get
accelerated in the dissipative fireballs. Interactions of protons
with fireball photons will produce a burst of neutrinos with
energies of $\sim$PeV (e.g. Waxman $\&$Bahcall 1997;  Guetta et al.
2004; Dermer $\&$ Atoyan 2006), so detection of such neutrinos would prove
the presence of cosmic ray protons in the fireball.  Despite that
large progresses in the studies of GRBs and their afterglow have
been made recently, the composition of the jet, whether it is
proton-electron dominated or Poynting-flux dominated, is largely
unknown. Both baryon-dominated  fireball shock model (e.g. Rees $\&$
M\'{e}sz\'{a}ros 1994; Paczy\'{n}ski $\&$ Xu 1994) and magnetic
dissipation model (e.g. Narayan $\&$ Kumar 2009; Zhang $\&$ Yan 2010)
have been proposed for the central engine of GRBs. In the
magnetically dominated outflow model for GRBs, the non-thermal proton
energy fraction may be low (but see Giannios 2010),
while in the baryon-dominated outflow model,
it is natural to expect proton acceleration via shock dissipation of the kinetic energy.
Since the flux of neutrinos depends on the
energy fraction of protons in the fireball for a given burst energy,
the flux or limit of the neutrino emission could constrain the
proton energy fraction and in principal provide a useful probe of
the jet composition. The proton energy fraction is also crucial to
know whether GRBs could provide sufficient flux for UHECRs.

The Kilometer-scale IceCube detector is the most sensitive neutrino
telescope in operation, although no positive neutrino signal has
been detected so far (Abbasi et al. 2010, 2011a;  The IceCube
collaboration 2011). The IceCube operations with the 22, 40 and 59
strings configurations all yield negative results, which have put
more and more stringent constraints on neutrino emission from GRBs.
The analysis were performed for both point source search from
individual GRBs and diffuse emission from aggregated GRBs (Abbasi et
al. 2011b). According to the IceCube collaboration, with 40-string
configuration operation between 2008 and 2009, IceCube reaches a
sensitivity at the level of the expect flux from GRBs and the
combined upper limit of IceCube 40-string and IceCube 59-string
analysis is 0.22 times the expected flux (Abbasi et al. 2011a).
Based on this, the IceCube collaboration argued that the UHECR-GRB
connection is challenged (Abbasi et al. 2011a).

The calculation by the IceCube collaboration (ICC hereafter) is
based on the formula in the appendix of the paper Abbasi et al.
(2010) and benchmark parameters for the internal shock model of
GRBs. However, as also pointed out by Li (2011) and H{\"u}mmer et
al. (2011),  we will show that the normalization procedure used by
ICC overestimates the neutrino flux. In calculating the photon
number density, ICC also approximates the energy of all photons by
the break energy of the photon spectrum, which originates from the
assumption in Guetta et al. 2004 (GT2004 hereafter). To get a more
accurate estimate of the expected neutrino flux, we first present a
refined analytic calculation in $\S$ 2.1, revising the above
approximations, and perform a numerical calculation in $\S$ 2.2
taking into account three main energy loss channels for protons
interacting with burst photons. In $\S$ 3, we study the effects of
non benchmark parameters on the neutrino emission, such as the
dissipation radius and the bulk Lorentz factor of the fireball. In
$\S$ 4, we calculate the accumulative diffuse neutrino emission from
GRBs and confront it with the IceCube limit on diffuse neutrinos. At
the end, we give our conclusions and discussions.

\section{GRB Neutrino Spectra}
Based on the assumption that protons and electrons are accelerated
in the same region of a GRB, protons interact with photons emitted
by electron synchrotron emission or inverse-Compton emission
predominantly produce the charged and neutral pions. The charged
pion subsequently decays to produce 4 final state leptons, via the
processes $\pi^{\pm} \rightarrow \nu_{\mu}(\bar{\nu}_\mu)\mu^{\pm}
 \rightarrow\nu_\mu(\bar{\nu}_\mu) e^+(e^-)\nu_e(\bar{\nu}_e)\bar{\nu}_\mu(\nu_\mu)$,
which approximately share the pion energy equally. Denoting $\Re$ as
the ratio between the charged pion  number to the total pion number,
the fraction of the proton energy lost into each lepton is
$\frac{\Re}{4}f_{p\gamma}$, where $f_{p\gamma}$ is the fraction of
the protons energy lost into pions. For the proton with energy of
$\epsilon_p=\gamma_p m_pc^2$, the photomeson interaction timescale
can be calculated by (Waxman $\&$ Bahcall 1997)
\begin{equation}\label{tpgreverse}
\begin{split}
t_{p\gamma}^{-1}(\epsilon_p)&=\frac{1}{\epsilon_p}\frac{d\epsilon_p}{dt}\\
&=\frac{c}{2\gamma_p^2}\int_{\tilde\epsilon_{{\gamma,\rm th}}}^\infty
d\tilde\epsilon_\gamma\sigma_{{p\gamma}}(\tilde\epsilon_\gamma)\xi(\tilde\epsilon_\gamma)\tilde\epsilon_\gamma
\int_{\tilde\epsilon_\gamma/2\gamma_p}^{\infty}dx
x^{-2}\frac{dn_\gamma}{dx},
\end{split}
\end{equation}
where $\sigma_{p\gamma}(\tilde\epsilon_\gamma)$ is the cross section of
photomeson interaction for a photon with energy $\tilde\epsilon_\gamma$ in
the proton-rest frame, $\xi(\tilde\epsilon_\gamma)$ is the
inelasticity,
$\tilde\epsilon_{\gamma,\rm th}$ is the threshold of photon energy and
$\frac{dn_\gamma}{d\epsilon_\gamma}$ is the GRB photon spectrum in
fluid-rest frame (Waxman $\&$ Bahcall 1997). The fraction of protons energy loss into pions is
\begin{equation}\label{fpgtheory}
 f_{p\gamma}
 =1-\exp(-t_{\rm dyn}/t_{p\gamma}),
\end{equation}
where
$t_{\rm dyn}=R/(\Gamma c)$ is the dynamic
timescale.
Just for simplicity, we use $t_{\rm dyn}$ as the interaction time. Generally speaking,
proton cooling timescales can be shorter than the dynamical timescale
(see Murase $\&$ Nagataki 2006a,b, for details), but we do not consider such complicated
effects for the purpose of testing the standard model suggested by Waxman $\&$ Bahcall (1997)
(Hummer et al. 2012).

Approximating that neutrinos produced via photomeson interaction
by proton with energy $\epsilon_p$ have constant energy,
the spectrum of neutrinos from the decay of secondary particles can
be obtained by, without considering the oscillation,
\begin{equation}\label{nunnupiontheory}
\epsilon_\ell\frac{dn_\ell}{d\epsilon_\ell}\mathrm{d}\epsilon_\ell
 = \frac{\Re(\epsilon_p)}{4}f_{p\gamma} (\epsilon_p)\theta_\ell(\epsilon_p )\epsilon_p \frac{dn_p}{d\epsilon_p}\mathrm{d}\epsilon_p  ,
\end{equation}
where $\frac{dn_p}{d\epsilon_p}$ is the spectrum of protons and $\Re$
is the ratio between the charged pion number to the total pion number as defined
above equation (\ref{tpgreverse}). The
subscript $\ell$ represents different flavors of neutrinos, i.e.,
$\ell=\nu_\mu, \bar\nu_\mu, \nu_e$ for muon neutrinos $\nu_\mu$
produced via the decay of pions, antimuon neutrinos $\bar\nu_\mu$ and
electron neutrinos $\nu_e$ produced via the decay of muons,
respectively. If the cooling timescale of pions or muons is smaller
than their lifetime, pions or muons have the probability to cool
down before decay.  Then the neutrino flux will be suppressed by a
factor of $\zeta_{\pi}=1-\exp{(-t_{\pi,\rm syn}/\tau_{\pi})}$,
where $t_{\pi,\rm syn}=3.3\times10^{-3}{\rm s}L_{\gamma,52}^{-1}\Gamma_{2.5}^2 R_{14}^2\epsilon_{\pi,\rm
EeV}^{-1} $ is the synchrotron cooling timescale,
 and $\tau_{\pi}=2.6\times 10^{-8}{\rm s}\gamma_{\pi}=186{\rm s}\epsilon_{\pi,\rm
EeV}$ is the lifetime of pions, whose energy is
$\epsilon_{\pi}=0.2\epsilon_{\rm p}$. Here we assume the fraction of
electron energy and magnetic field energy are same, i.e.,
$\epsilon_e=\epsilon_B$. Similarly, the suppression due to muon
cooling is $\zeta_{\mu}=1-\exp{(-t_{\mu,\rm syn}/\tau_{\mu})}$,
where $t_{\mu,\rm syn}=1.1\times10^{-3}{\rm
s}L_{\gamma,52}^{-1}\Gamma_{2.5}^2 R_{14}^2\epsilon_{\mu,\rm
EeV}^{-1}$ and $\tau_{\mu}=2.1\times 10^4{\rm s}\epsilon_{\mu,\rm
EeV}$ with the muon energy $\epsilon_{\mu}=0.15\epsilon_{\rm p}$.
The suppression factor $\theta_\ell(\epsilon_p )$ is a combination of
$\zeta_{\pi}(\epsilon_p)$ and $\zeta_\mu(\epsilon_p)$, i.e.,
$\theta_{\nu_{\mu}}(\epsilon_p )=\zeta_{\pi}(\epsilon_p)$ and
$\theta_{\bar\nu_{\mu}(\nu_e)}(\epsilon_p )=\zeta_\pi(\epsilon_p)\zeta_\mu(\epsilon_p)$.

Considering the neutrino oscillation effect, the spectrum  of muon
neutrinos (including antimuon neutrinos) detected on Earth is
approximated as ( Nagataki et al. 2003; Particle Data Group 2004; Kashti $\&$ Waxman 2005;
Murase 2007; Li 2011)
\begin{equation}\label{oscillation}
\frac{dn_\nu}{d\epsilon_\nu}=0.2\frac{dn_{\nu_e}}{d\epsilon_{\nu_e}}
+0.4\frac{dn_{\nu_\mu}}{d\epsilon_{\nu_\mu}}+0.4\frac{dn_{\bar\nu_\mu}}{d\epsilon_{\bar\nu_\mu}},
\end{equation}
where $\nu_\mu$ are produced via the decay of secondary pions, and
$\nu_e$ and $\bar\nu_\mu$ are produced via the decay of secondary
muons. Therefore, by inserting equation (\ref{nunnupiontheory}) into
equation (\ref{oscillation}), one can calculate the total spectrum
of muon (including antimuon) neutrinos detected on Earth numerically
via the following equation:
\begin{equation}\label{nunnutotaltheory}
\epsilon_{\nu} \frac{dn_{\nu}}{d\epsilon_{\nu}}\mathrm{d}\epsilon_{\nu}
= \frac{\Re(\epsilon_p)}{4}f_{p\gamma}(\epsilon_p)\theta_\nu(\epsilon_p)
 \epsilon_p \frac{dn_p}{d\epsilon_p}\mathrm{d}\epsilon_p ,
\end{equation}
with the  factor
$\theta_\nu(\epsilon_p)=0.4\zeta_\pi(\epsilon_p)+0.6\zeta_\pi(\epsilon_p)\zeta_\mu(\epsilon_p)$
accounting for the neutrino oscillation and the cooling of secondary
particles.

The GRB photon distribution can be described by
\begin{equation}\label{ngamma}
\frac{dn_\gamma}{d\epsilon_\gamma}=A_\gamma
\left(\frac{\epsilon_\gamma}{\epsilon_{\gamma\rm b}}\right)^{-q}
\end{equation}
where $\epsilon_{\gamma, b}$ is the break energy (in the fluid-rest
frame) of the photon spectrum, $q=\alpha$ for
$\epsilon_{\gamma}<\epsilon_{\gamma b}$ and $q=\beta$ for
$\epsilon_{\gamma}>\epsilon_{\gamma \rm b}$. The normalized
coefficient is
$A_\gamma=U_\gamma[\int_{\epsilon_{\gamma,\rm
min}}^{\epsilon_{\gamma,\rm max}}\epsilon_\gamma
\left(\frac{\epsilon_\gamma}{\epsilon_{\gamma\rm
b}}\right)^{-q}d\epsilon_\gamma]^{-1}
=\frac{U_\gamma}{y_1\epsilon_{\gamma\rm b}^2}$,
where the energy density of photons is
$U_{\gamma}=\frac{L_{\gamma}}{4\pi R^2\Gamma^2c}$,
and
\begin{equation}\label{y1}
y_1=\frac{1}{\alpha-2}\left(\frac{\epsilon_{\gamma\rm b}}{\epsilon_{\gamma,\rm min}}\right)^{\alpha-2}
-\frac{1}{\beta-2}\left(\frac{\epsilon_{\gamma\rm b}}{\epsilon_{\gamma,\rm max}}\right)^{\beta-2}-\frac{1}{\alpha-2}+\frac{1}{\beta-2}
\end{equation}
for $\beta\neq2$, where $\epsilon_{\gamma,\rm min}$ and
$\epsilon_{\gamma,\rm max}$ are the minimum and maximum energy of
the photons. For $\beta=2$,
$y_1=\left(1-\left(\frac{\epsilon_{\gamma,\rm
min}}{\epsilon_{\gamma\rm b}}\right)^{-\alpha+2}\right)/(-\alpha+2)
+\ln\left(\frac{\epsilon_{\gamma,\rm max}}{\epsilon_{\gamma\rm
b}}\right)$. Hereafter, we adopt assumptions that
$\epsilon_{\gamma,\rm min}=1{\rm keV}$ and $\epsilon_{\gamma,\rm
max}=10\rm MeV$ as ICC did (Abassi et al. 2010). If we know redshift
of bursts, the correction should be properly taken into account in
estimating the luminosity.

The  proton number per energy interval can be described by
$\frac{dn_p}{d\epsilon_p}=N_p\epsilon_p^{-s}$ with $s$ being the
power-law index and $N_p$ being the normalized coefficient. The
normalized coefficient of the injected proton spectrum can be
calculated by $N_p=E_p/\int_{\epsilon_{\rm p,min}}^{\epsilon_{\rm
p,max}} d\epsilon_p \epsilon_p^{1-s} =E_p/\ln{\frac{\epsilon_{\rm
p,max}}{\epsilon_{\rm p,min}}}$ (we set $s= 2$ hereafter, as
predicted by the shock acceleration theory) with $E_p$ being the
total energy in protons, $\epsilon_{\rm p,min}$ and $\epsilon_{\rm
p,max}$ being the minimum and maximum  energy of accelerated
protons, respectively. We introduce a factor $\eta_p$,  denoting the
ratio of the energy in accelerated protons to the radiation energy,
then the total proton energy $E_p$ is (Murase $\&$ Nagataki 2006a)
\begin{equation}
E_p=\eta_p E_{\rm iso}
\end{equation}
where $E_{\rm iso}$ is the isotropic energy of the burst, which is
obtained  from the observed fluence $F_\gamma^{\rm ob}$ in the
energy band of 1 keV - 10 MeV and the redshift of the burst.
According to Waxman (1995) and Li (2011), for mildly-relativistic
GRB internal shocks, we assume the minimum proton energy as $\epsilon_{\rm
p,min}^{\rm ob}\simeq \Gamma
m_pc^2=3.0\times10^{11}\Gamma_{2.5}/(1+z)\rm eV$ \footnote{ In some
other papers, $\epsilon_{\rm p,min}^{\rm ob}\simeq 4\Gamma m_pc^2$
or $\epsilon_{\rm p,min}^{\rm ob}\simeq 10\Gamma m_pc^2$ are
adopted, since the relative Lorentz factor is order of 1-10, where
higher values favor efficiency internal shocks.} , and the maximum
proton energy due to synchrotron cooling is $\epsilon_{\rm
p,max}^{\rm
ob}=4.0\times10^{20}\Gamma_{2.5}^{3/2}R_{14}^{1/2}\epsilon_e^{1/4}
\epsilon_{B}^{-1/4}g_1^{-1/2}L_{\gamma,52}^{-1/4}/(1+z){\rm eV}$,
with $g_1\ga 1$ being a factor accounting for the uncertainty in the
particle acceleration time.

\subsection{Analytical Calculation}
\subsubsection{Neutrino spectrum in the general dissipation scenario}
In this subsection, we treat the dissipation radius as a free
parameter, since the exact dissipation mechanism of GRBs is not
established. There are suggestions that, besides the standard
internal shock model, the prompt emission arises from the
dissipative photosphere or arises at much larger radii where the
magnetic-dominated outflow is dissipated through magnetic dissipation processes,
such as reconnection (e.g. Narayan $\&$ Kumar 2009, Kumar $\&$
Narayan 2009, Zhang $\&$ Yan 2010).

For the analytical calculation, we adopt the $\Delta$ resonance
approximation  as in Waxman $\&$ Bahcall (1997) and Guetta et al.
(2004), where the cross section peaks at the photon energy
$\tilde\epsilon_\gamma\sim\epsilon_{\rm peak}=0.3{\rm GeV}$ in the
proton-rest frame. If $t_{\rm dyn}< t_{p\gamma}$, the conversion
fraction is approximated as $f_{p\gamma}\simeq t_{\rm
dyn}/t_{p\gamma}=R/(\Gamma c t_{p\gamma})$. Adopting the $\Delta$
resonance approximation (Waxman $\&$ Bahcall 1997), the fraction of
proton energy converted into pion is
\begin{eqnarray}\label{fpgR}
f_{p\gamma}(\epsilon_{\rm p}^{\rm ob})&\simeq&\frac{0.11}{y_1}\left(\frac{2}{\alpha+1}\right)\left(\frac{1}{1+z}\right)
\frac{L_{\gamma 52}}{\epsilon_{\gamma b,\rm MeV}^{\rm ob}\Gamma_{2.5}^2R_{14}}\nonumber\\
&&\times
\begin{cases}
k_1\left(\frac{\epsilon_{p}^{\rm ob}}{\epsilon_{\rm p,b}^{\rm ob}}\right)^{\beta-1},
\,\,\,\,\,\,\,\,\,\,\,\epsilon_{p}^{\rm ob}\le\epsilon_{\rm p,b}^{\rm ob}\\
\left(\frac{\epsilon_{p}^{\rm ob}}{\epsilon_{\rm p,b}^{\rm ob}}\right)^{\alpha-1}
+k_{p},
\,\,\,\,\,\epsilon_{p}^{\rm ob}>\epsilon_{\rm p,b}^{\rm ob}\\
\end{cases}
\end{eqnarray}
where
\begin{equation}\label{epb}
\epsilon_{\rm p,b}^{\rm ob}=\Gamma_{\rm p}^{\rm ob}m_pc^2
=\frac{\Gamma^2\xi_{\rm peak}}{2(1+z)^2\epsilon_{\gamma b}^{\rm ob}},
\end{equation}
$k_1=\frac{\alpha+1}{\beta+1}$ and
$k_p=\frac{\alpha-\beta}{\beta+1}\left(\frac{\epsilon_p^{\rm
ob}}{\epsilon_{\rm p,b}^{\rm ob}}\right)^{-2}$ \footnote{ Hereafter,
for brevity, we abandoned the coefficient $k_1$ and $k_\ell$ in the
following equations, since $k_1\simeq 1$ and $k_\ell\simeq 0$
approximately, where
$k_\ell=\frac{\alpha-\beta}{\beta+1}\left(\frac{\epsilon_\ell^{\rm
ob}}{\epsilon_{\ell,\rm b}^{\rm ob}}\right)^{-2}$ with $\ell$
represents three flavors of neutrinos, i.e., electron neutrinos and
muon(antimuon) neutrinos. But we still adopt that in our analytic
calculations.}. Note that the above approximation is valid when the
radius is not too small, i.e., $R>
1.1\times10^{13}\left(\frac{1}{y_1}\right)\left(\frac{2}{\alpha+1}\right)\left(\frac{1}{1+z}\right)
L_{\gamma, 52}\epsilon_{\gamma b,\rm MeV}^{\rm
ob,-1}\Gamma_{2.5}^{-2}{\rm cm}$.

As $\epsilon_{\nu}=0.05\epsilon_{p}$ for the $\Delta$ resonance
approximation, from equation (\ref{epb}), we can obtain the break
energy of neutrino spectrum corresponding to the photon spectral
break
\begin{equation}\label{enp}
\epsilon_{\nu,b}^{\rm ob}=7.5\times10^{5}{\rm GeV}(1+z)^{-2}
\Gamma_{2.5}^2\epsilon_{\gamma,\rm MeV}^{\rm ob,-1}.
\end{equation}
The cutoff energy of muon neutrino spectrum due to the pion cooling
can be obtained by setting $t_{\pi,\rm syn}=\tau_\pi$,
\begin{equation}\label{encpionR}
\epsilon_{\nu_\mu,\rm c}^{\rm ob}=\frac{3.3\times10^8}{(1+z)}L_{\gamma,52}^{-1/2}\Gamma_{2.5}^2R_{14}{\rm GeV}.
\end{equation}
For antimuon neutrinos (and electron neutrinos) produced via the
decay of muons, an extra break is caused by  the muon cooling, which
is
\begin{equation}\label{encmuonR}
\epsilon_{\lambda,\rm c}^{\rm ob}=\frac{2.4\times10^7}{(1+z)}L_{\gamma,52}^{-1/2}\Gamma_{2.5}^2R_{14}{\rm GeV},
\end{equation}
where the subscript $\lambda$ represents antimuon neutrino
$\bar\nu_\mu$ or electron neutrino $\nu_e$.

Assuming that the fraction of the amount of charged pions is
$\Re=1/2$, from equations (\ref{nunnupiontheory}) and (\ref{fpgR}),
we get the spectrum of muon neutrinos
produced by the pion decay,
\begin{eqnarray}\label{nunnupionR}
&&\,\,\,\,\,\,\,(\epsilon_{\nu_\mu}^{\rm ob})^{2}\frac{dn_{\nu_\mu}}{d\epsilon_{\nu_\mu}^{\rm ob}}\nonumber\\
&&=\frac{0.014}{y_1}\left(\frac{2}{\alpha+1}\right)\left(\frac{1}{1+z}\right)
\frac{\eta_pF_{\gamma}^{\rm ob}}{\ln(\frac{\epsilon_{\rm p,max}^{\rm ob}}{\epsilon_{\rm p,min}^{\rm ob}})}
\frac{L_{\gamma, 52}}{\epsilon_{\gamma b,\rm MeV}^{\rm ob}\Gamma_{2.5}^2R_{14}}\nonumber\\
&&\times
\begin{cases}
\left(\frac{\epsilon_{{\nu_\mu}}^{\rm ob}}{\epsilon_{{\nu},b}^{\rm ob}}\right)^{\beta-1},
\,\,\,\,\,\,\,\,\,\,\,\,\,\,\,\,\,\,\,\,\,\,\,\,\,\,\,\,\,\epsilon_{{\nu_\mu}}^{\rm ob}\le\epsilon_{{\nu},b}^{\rm ob}\\
\left(\frac{\epsilon_{{\nu_\mu}}^{\rm ob}}{\epsilon_{{\nu},b}^{\rm ob}}\right)^{\alpha-1},
\,\,\,\,\,\,\,\,\,\,\,\epsilon_{{\nu},b}^{\rm ob}<\epsilon_{{\nu_\mu}}^{\rm ob}\le\epsilon_{{\nu_\mu},\rm c}^{\rm ob}\\
\left(\frac{\epsilon_{{\nu_\mu}}^{\rm ob}}{\epsilon_{{\nu},b}^{\rm ob}}\right)^{\alpha-1}
\left(\frac{\epsilon_{\nu_\mu}^{\rm ob}}{\epsilon_{{\nu_\mu},\rm c}^{\rm ob}}\right)^{-2}.
\,\,\,\,\epsilon_{{\nu_\mu}}^{\rm ob}>\epsilon_{{\nu_\mu},\rm c}^{\rm ob}\\
\end{cases}
\end{eqnarray}
Similarly, the spectrum of antimuon (and electron) neutrinos
produced by muon decay is approximated by
\begin{eqnarray}\label{nunnumuonR}
&&\,\,\,\,\,\,\,(\epsilon_\lambda^{\rm ob})^{2}\frac{dn_\lambda}{d\epsilon_\lambda^{\rm ob}}\nonumber\\
&&=\frac{0.014}{y_1}\left(\frac{2}{\alpha+1}\right)\left(\frac{1}{1+z}\right)\frac{\eta_pF_{\gamma}^{\rm ob}}{\ln(\frac{\epsilon_{\rm p,max}^{\rm ob}}{\epsilon_{\rm p,min}^{\rm ob}})}
\frac{L_{\gamma 52}}{\epsilon_{\gamma b,\rm MeV}^{\rm ob}\Gamma_{2.5}^2R_{14}}\nonumber\\
&&\times
\begin{cases}
\left(\frac{\epsilon_\lambda^{\rm ob}}{\epsilon_{\nu,b}^{\rm ob}}\right)^{\beta-1},
\,\,\,\,\,\,\,\,\,\,\,\,\,\,\,\,\,\,\,\,\,\,\,\,\,\,\,\,\,\,\,\,\,\,\,\,\,\,\,\,\,\,\,\,\,\,
\epsilon_\lambda^{\rm ob}\le\epsilon_{\nu,b}^{\rm ob}\\
\left(\frac{\epsilon_\lambda^{\rm ob}}{\epsilon_{\nu,b}^{\rm ob}}\right)^{\alpha-1},
\,\,\,\,\,\,\,\,\,\,\,\,\,\,\,\,\,\,\,\,\,\,\,\,\,\,\,\,\,\,\,\,
\epsilon_{\nu,b}^{\rm ob}<\epsilon_\lambda^{ob}\le\epsilon_{\lambda,\rm c}^{\rm ob}\\
\left(\frac{\epsilon_\lambda^{\rm ob}}{\epsilon_{\nu,b}^{\rm ob}}\right)^{\alpha-1}
\left(\frac{\epsilon_\lambda^{ob}}{\epsilon_{\lambda,\rm c}^{ob}}\right)^{-2},
\,\,\,\,\,\,\,\,
\epsilon_{\lambda,\rm c}^{\rm ob}<\epsilon_\lambda^{ob}<\epsilon_{{\nu_\mu},\rm c}^{\rm ob}\\
\left(\frac{\epsilon_\lambda^{\rm ob}}{\epsilon_{\nu,b}^{\rm ob}}\right)^{\alpha-1}
\left(\frac{\epsilon_\lambda^{\rm ob}}{\epsilon_{\lambda,\rm c}^{\rm ob}}\right)^{-2}
\left(\frac{\epsilon_\lambda^{\rm ob}}{\epsilon_{{\nu_\mu},\rm c}^{ob}}\right)^{-2},
\epsilon_{\nu_\mu,\rm c}^{\rm ob}<\epsilon_\lambda^{ob}\\
\end{cases}
\end{eqnarray}
.Then, one can
obtain the $\nu_\mu+\bar{\nu}_\mu$ spectrum after considering the
neutrino oscillation effect by substituting equations
(\ref{nunnupionR}) and (\ref{nunnumuonR}) into equation
(\ref{oscillation}).

\subsubsection{Neutrino spectrum in the internal shock scenario}
In this subsection, we assume the standard internal shock scenario
with the dissipation radius at $R=2\Gamma^2ct_{\rm v}^{\rm ob}/(1+z)$, where
$t_{\rm v}^{\rm ob}$ is the observed variability timescale of GRB emission.
The conversion fraction $f_{p\gamma}$ is
given by
\begin{eqnarray}\label{fpg}
f_{p\gamma}(\epsilon_{\rm p}^{\rm ob})&\simeq&\frac{0.18}{y_1}\left(\frac{2}{\alpha+1}\right)
\frac{L_{\gamma 52}}{\epsilon_{\gamma b,\rm MeV}^{\rm ob}\Gamma_{2.5}^4t_{\rm v,-2}^{\rm ob}}\nonumber\\
&&\times
\begin{cases}
\left(\frac{\epsilon_{p}^{\rm ob}}{\epsilon_{\rm p,b}^{\rm ob}}\right)^{\beta-1},
\,\,\epsilon_{p}\le\epsilon_{\rm p,b}^{\rm ob}\\
\left(\frac{\epsilon_{p}^{\rm ob}}{\epsilon_{\rm p,b}^{\rm
ob}}\right)^{\alpha-1}.
\,\,\,\,\epsilon_{p}>\epsilon_{\rm p,b}^{\rm ob}\\
\end{cases}
\end{eqnarray}
Then the spectrum of muon neutrinos produced via pion decay is approximated by
\begin{eqnarray}\label{nunnupion}
(\epsilon_{\nu_\mu}^{\rm ob})^{2}\frac{dn_{\nu_\mu}}{d\epsilon_{\nu_\mu}^{\rm ob}}
&=&\frac{0.023}{y_1}\left(\frac{2}{\alpha+1}\right)\frac{\eta_pF_{\gamma}^{\rm ob}}{\ln(\frac{\epsilon_{\rm p,max}^{\rm ob}}{\epsilon_{\rm p,min}^{\rm ob}})}
\frac{L_{\gamma, 52}}{\epsilon_{\gamma b,\rm MeV}^{\rm ob}\Gamma_{2.5}^4t_{\rm v,-2}^{\rm ob}}\nonumber\\
&&\times
\begin{cases}
\left(\frac{\epsilon_{{\nu_\mu}}^{\rm ob}}{\epsilon_{{\nu},b}^{\rm ob}}\right)^{\beta-1},
\,\,\,\,\,\,\,\,\,\,\,\,\,\,\,\,\,\,\,\,\,\,\,\,\,\,\,\,\,\,\epsilon_{{\nu_\mu}}^{\rm ob}\le\epsilon_{{\nu},b}^{\rm ob}\\
\left(\frac{\epsilon_{{\nu_\mu}}^{\rm ob}}{\epsilon_{{\nu},b}^{\rm ob}}\right)^{\alpha-1},
\,\,\,\,\,\,\,\,\,\,\,\,
\epsilon_{{\nu},b}^{\rm ob}<\epsilon_{{\nu_\mu}}^{\rm ob}\le\epsilon_{{\nu_\mu},\rm c}^{\rm ob}\\
\left(\frac{\epsilon_{{\nu_\mu}}^{\rm ob}}{\epsilon_{{\nu},\rm b}^{\rm ob}}\right)^{\alpha-1}
\left(\frac{\epsilon_{\nu_\mu}^{\rm ob}}{\epsilon_{{\nu_\mu},\rm c}^{\rm ob}}\right)^{-2},
\,\,\,\epsilon_{{\nu_\mu}}^{\rm ob}>\epsilon_{{\nu_\mu},\rm c}^{\rm ob}\\
\end{cases}
\end{eqnarray}
and the spectrum of antimuon(electron) neutrinos produced via muon decay is approximated by
\begin{eqnarray}\label{nunnumuon}
(\epsilon_\lambda^{\rm ob})^{2}\frac{dn_\lambda}{d\epsilon_\lambda^{\rm ob}}
&=&\frac{0.023}{y_1}\left(\frac{2}{\alpha+1}\right)\frac{\eta_pF_{\gamma}^{\rm ob}}{\ln(\frac{\epsilon_{\rm p,max}^{\rm ob}}{\epsilon_{\rm p,min}^{\rm ob}})}
\frac{L_{\gamma, 52}}{\epsilon_{\gamma b,\rm MeV}^{\rm ob}\Gamma_{2.5}^4t_{\rm v,-2}^{\rm ob}}\nonumber\\
&&\times
\begin{cases}
\left(\frac{\epsilon_\lambda^{\rm ob}}{\epsilon_{\nu,b}^{\rm ob}}\right)^{\beta-1},
\,\,\,\,\,\,\,\,\,\,\,\,\,\,\,\,\,\,\,\,\,\,\,\,\,\,\,\,\,\,\,\,\,\,\,\,\,\,\,\,\,\,\,\,\,\,
\epsilon_\lambda^{\rm ob}\le\epsilon_{\nu,b}^{\rm ob}\\
\left(\frac{\epsilon_\lambda^{\rm ob}}{\epsilon_{\nu,\rm b}^{\rm ob}}\right)^{\alpha-1},
\,\,\,\,\,\,\,\,\,\,\,\,\,\,\,\,\,\,\,\,\,\,\,\,\,\,\,\,\,\,\,\,
\epsilon_{\nu,\rm b}^{\rm ob}<\epsilon_\lambda^{\rm ob}\le\epsilon_{\lambda,\rm c}^{\rm ob}\\
\left(\frac{\epsilon_\lambda^{\rm ob}}{\epsilon_{\nu,\rm b}^{\rm ob}}\right)^{\alpha-1}
\left(\frac{\epsilon_\lambda^{\rm ob}}{\epsilon_{\lambda,\rm c}^{\rm ob}}\right)^{-2},
\,\,\,\,\,\,\,\,
\epsilon_{\lambda,\rm c}^{\rm ob}<\epsilon_\lambda^{\rm ob}<\epsilon_{{\nu_\mu},\rm c}^{\rm ob}\\
\left(\frac{\epsilon_\lambda^{\rm ob}}{\epsilon_{\nu,\rm b}^{\rm ob}}\right)^{\alpha-1}
\left(\frac{\epsilon_\lambda^{\rm ob}}{\epsilon_{\lambda,\rm c}^{\rm ob}}\right)^{-2}
\left(\frac{\epsilon_\lambda^{\rm ob}}{\epsilon_{{\nu_\mu},\rm c}^{\rm ob}}\right)^{-2},
\epsilon_{\nu_\mu,\rm c}^{\rm ob}<\epsilon_\lambda^{\rm ob}\\
\end{cases}
\end{eqnarray}
where the cutoff energies are
\begin{equation}\label{encpion}
\epsilon_{\nu_\mu,\rm c}^{\rm ob}=\frac{2.0\times10^8}{(1+z)^2}L_{\gamma,52}^{-1/2}\Gamma_{2.5}^4t_{\rm v,-2}^{\rm ob}{\rm GeV},
\end{equation}
and
 \begin{equation}\label{encmuon}
\epsilon_{\lambda,\rm c}^{\rm
ob}=\frac{1.4\times10^7}{(1+z)^2}L_{\gamma,52}^{-1/2}\Gamma_{2.5}^4t_{\rm
v,-2}^{\rm ob}{\rm GeV},
\end{equation}
with $\lambda$ representing antimuon and electron neutrinos
($\bar\nu_{\mu}$ and $\nu_e$) produced by muon decay, and
\begin{equation}\label{epmax}
\frac{\epsilon_{\rm p, max}^{\rm ob}}{\epsilon_{\rm p,min}^{\rm ob}}=1.0\times10^{9}
\Gamma_{2.5}^{3/2}(t_{\rm v,-2}^{\rm ob})^{1/2}L_{\gamma,52}^{-1/4}
\epsilon_e^{1/4}
\epsilon_{B}^{-1/4}g_1^{-1/2}.
\end{equation}

Substituting equations (\ref{nunnupion}) and (\ref{nunnumuon}) into
equation (\ref{oscillation}), we can obtain the neutrino
spectrum analytically. To illustrate the difference between our
calculation and the ICC calculation, we calculate the neutrino
spectrum for one typical GRB with benchmark parameters, shown in
Figure 1.  Compared with the ICC calculation (the dark gray solid line),
our spectrum (the purple solid line) consists of more structures
resulting from the sum of the contributions by three flavors of
neutrinos, for which both pion cooling,  muon cooling and the
oscillation effect are considered. Furthermore, the flux level
predicted by our modified analytical calculation is a factor of
$\sim 20$ lower than that obtained by ICC (Abbasi et al., 2011).
This mainly arises from two differences in the calculation:

(I) We use equation (\ref{nunnupiontheory}), where the conversion
fraction $f_{p\gamma}$ is a function of the proton energy
$\epsilon_{p}$ as shown by equation (\ref{fpg}), to normalize the
neutrino flux to the proton flux, which means that only a fraction
of protons can produce neutrinos efficiently. This corrects ICC's
inaccurate use of energy-independent conversion fraction in the
normalization of the neutrino flux (Li 2011; H{\"u}mmer et al. 2011;
Murase et al. 2012). The calculation of Guetta et al. (2004)
\footnote{Guetta et al. (2004) calculated the neutrino spectrum by
assuming a flat high-energy electron spectrum (i.e.
$dN_e/d\gamma_e\propto \gamma_e^{-2}$) and using an electron
equipartition fraction $\epsilon_e^G$ that represents the ratio of
the nonthermal electron energy over one energy decade to the UHECR
energy over one energy decade, and the neutrino flux is normalized
by $\epsilon_\nu^2 dN_\nu/d\epsilon_\nu = (1/8) (1/\epsilon_e^G)
(F_\gamma^{ob} /{\rm ln10}) f_{\pi}$ (see their Eq. A19). Note that
other normalization procedures are also possible, and this
$\epsilon_e^G$ is typically larger than the conventional
$\epsilon_e$ that is defined as the ratio of the total nonthermal
electron energy to the total internal energy (including both thermal
and nonthermal protons).} normalized the flux based on the
differential spectrum so that it does not suffer from this problem.
The spectrum obtained with the calculation of Guetta et al.
(2004)\footnote{In our paper, we used the nonthermal baryon-loading
parameter $\eta_p$, defined as the ratio between the total energy in
accelerating protons and the total radiation energy at observed
bands, to normalize the neutrino flux through $\epsilon_\nu^2
dN_\nu/d\epsilon_\nu = (1/8) (\eta_p/{\rm ln}(\epsilon_{p,\rm
max}/\epsilon_{p, \rm min})) F_\gamma^{ob}f_{p\gamma}$. The lines of
``modified Guetta  et al. (2004) " in Figures 1 and 2 are obtained
by using the benchmark value of $\eta_p=10$, which corresponds to
$\epsilon_e^G={\rm ln}(\epsilon_{p,\rm max}/\epsilon_{p,\rm
min})/{\rm ln}(10) (1/\eta_p)\sim 1$ in Eq.(A19) in Guetta et al.
(2004). Choosing $\epsilon_e^G=0.1$ in Guetta et al. (2004) would
correspond to a higher baryon loading factor of $\eta_p={\rm
ln}(\epsilon_{p,\rm max}/\epsilon_{p,\rm min})/{\rm
ln}(10)(1/\epsilon_e^G) \sim 100$ in our paper. Note that the
GRB-UHECR hypothesis might suggest such high loading factors if the
local GRB rate is $0.01-0.1 \rm Gpc^{-3} yr{^-1}$ or the local
luminosity density is significantly smaller than $\sim10^{44}\rm erg
Mpc^{-3} yr^{-1}$, and those relatively optimistic cases are
constrained by IceCube under reasonable assumptions, as expected in
Murase $\&$ Nagataki (2006b)."}
 is shown by the blue solid line in Figure 1, where
the middle of equation (A.15) in their paper is used assuming the
bolometric luminosity as the luminosity at the break energy. The
flux is lower than the ICC result (the dark gray solid line in
Figure 1) by a factor of $\sim4$.

(II) In calculating the photon number density, we consider the
photon energy distribution according to the real photon spectrum,
reflected by the normalized coefficient
$A_\gamma\simeq\frac{U_\gamma}{y_{1}\epsilon_{\gamma\rm b}^2}$,
where $y_1$ is shown in equation (\ref{y1}), while ICC approximate
the energy of all  photons by the break energy of the photon
spectrum. This leads to a flux a factor of $\sim$3-6 lower than the ICC
result for typical $\alpha$ and $\beta$ values and
$\epsilon_{\gamma b}\sim100-1000{\rm keV}$ in the GRB spectrum.

Finally, we get a neutrino spectral flux (purple solid line) lower than the one
predicted by ICC (dark gray solid line) by a factor of $\sim 20$ for a typical GRB, as shown in Figure 1. Of
course, the suppression factor is different for GRBs with different
parameters.

\subsection{Numerical Results}
Besides the baryon resonance, the direct pion process, multi-pion
process and the diffractive scattering also contribute to the total
$p\gamma$ cross section. Therefore, in our numerical calculation, we
adopt a more precise cross section for photomeson interaction,
including three main channels, i.e., the $\Delta$ resonance process,
direct pion process and multi-pion process. For simplicity, we assume that the
inelasticity is $\xi=0.2$ for $\bar\epsilon_\gamma<983{\rm MeV}$ and
$\xi=0.6$ for $\bar\epsilon_\gamma \ga 983{\rm MeV}$(Atoyan $\&$ Dermer
2001). Then, inserting the cross sections and inelasticity into equation
(\ref{tpgreverse}) and using equation (\ref{fpgtheory}), we can get
the fractions of protons energy converted to pions. The average
fractions of the charged pions are set to $\Re_{\Delta}=1/3$
for $\Delta$ Resonance process, $\Re_{\rm dir}=2/3$ for direct pion
production, and $\Re_{\rm mul}=2/3$ for multipion production,
based on numerical investigations (Mucke et al. 1999, 2000, Murase $\&$ Nagataki 2006a,
b, Murase et al., 2006, Murase, 2007, Baerwald et al. 2011).
In addition, we assume the neutrino energy is $\epsilon_\nu=0.05\epsilon_p$ for
$\Delta$-resonance and direct pion production channels, and
$\epsilon_\nu=0.03\epsilon_p$ for multipion production channel.
Inserting these quantities into equations (\ref{tpgreverse}),
(\ref{fpgtheory}) and then (\ref{nunnutotaltheory}), we can obtain
the spectrum of neutrino emission produced via the three dominant
channels. Although this simplified numerical approach is different from more
detailed, fully numerical calculations (Murase $\&$ Nagataki 2006a,b; Murase 2007;
Baerward et al. 2011), it is enough for our purpose and saves calculation time.

In Figure 1, we also show this numerical result (the red solid line)
for comparison. The flux obtained with this numerical calculation is
about 2-3 times larger than the analytical result in the energy range
$10^5{\rm GeV}$--$3\times10^{6}$ GeV for typical GRB , consistent
with Murase $\&$ Nagataki (2006a, b) and Baerwald et al. (2011).
So the analytic calculation presented in the above section
can still be used as a rough approximation.

\subsection{Confronting the calculations with IC59+40 observations}
215 GRBs are observed during the operations in the 40-string
and 59-string configurations of IceCube, yielding negative results. In this
section, we calculate the neutrino flux for the same 215 GRBs using
the same burst parameters as ICC.  The information for these samples
was taken from the website grbweb.icecube.wisc.edu\footnote{For some
information from this website with typo, which is different from
GCN, we take the information from GCN.} and GCN. For the unmeasured
parameters, we adopt the same average values as in Abbasi et al.
(2010, 2011a). We assume a ratio of proton energy to radiation
energy $\eta_p=10$, equivalent to $1/f_{e}=10$ in ICC
calculation. We adopt the internal shock model with shock radius at
$R=2\Gamma^2 c t_{\rm v}^{\rm ob}/(1+z)$, where the Lorentz factor is
$\Gamma=10^{2.5}$ and the observed variability timescale is $t_{\rm
v}^{\rm ob}=0.01{\rm s}$ for each long GRB and $t_{\rm v}^{\rm ob}=0.001{\rm s}$
for each short GRB as the ICC did. The
diffuse neutrino flux can be obtained by the total neutrino fluence for 215 individual GRBs
timing a factor $7.83\times 10^{-9} \rm sr^{-1} s^{-1}$
 with the assumption that a total of 667
uniformly GRBs are generated per year. By adopting the effective area of
IC59 and IC40 as a function of zenith angle, we can calculate the
expected number of neutrinos with energy from $10^5$ GeV to
$3\times10^6$ GeV (Abbasi et al. 2010, 2011a).
The corresponding combined 90$\%$ confidence level (CL) upper limit spectrum can be
obtained by assuming that  the limit amount of neutrino events is $n_{\rm lim}\simeq1.9$
in the energy range from $10^5$ GeV to $3\times10^6$ GeV
as in the IceCube collaboration (2011).

In Figure 2, we show the 215 neutrino spectra (light red thin solid
lines) for individual GRBs. The sum of them is presented as the red
thick solid line, which is about a factor of $\sim 10$ lower than
that predicted by ICC (the dark gray solid line; The IceCube
collaboration 2011).
As we can see, the neutrinos predicted by ICC2010 and GT2004 (dark gray
solid line and blue solid line) are above the 90$\%$ CL upper limits (dark gray
dashed line and blue dashed line) for the combined IC40 and IC59 data analysis,
while our predicted neutrino flux is below the
corresponding upper limit (the red dashed line). We find that the
total expected number of neutrinos with energy from $10^5$ GeV to
$3\times10^6$ GeV is 0.74, which is $36\%$ of the 90$\%$ CL upper
limit.

\section{Constraints on the fireball properties }

\subsection{Uncertainty in the dissipation radius}
In $\S$ 2.1.1, we  do not assume a specific dissipation model for
GRB emission, but leave the dissipation radius as a free parameter.
Internal shocks happen over a wide parameter range
(e.g. Nakar $\&$ Piran 2002), and even larger radii are suggested in some
other dissipation models (e.g.,
 Narayan $\&$ Kumar 2009, Zhang $\&$ Yan 2010).
In Figure 3, we show the total neutrino spectra for 215 GRBs by
assuming some fixed dissipation radii for every GRBs in the range
$R=10^{12}{\rm cm}-10^{16}{\rm cm}$ \footnote{We note that, if the
radius is smaller than the photosphere radius, the neutrino emission
produced by $p-p$ interactions becomes important (Wang $\&$ Dai
2009; Murase 2008), which is not considered here.}. It shows that
the neutrino flux for the case of $R=10^{12}{\rm cm}$ (the black
solid line) would exceed the corresponding IceCube upper limit (the
black dashed line) as long as the baryon loading factor is sufficiently greater
than unity. If we fix $\eta_{p}=10$, the non-detection
requires that the dissipation radius should be larger than $4\times
10^{12}{\rm cm}$. We note that, when the emission radius is too
small, the maximum energy of accelerating particles is limited due
to strong photohadronic and/or radiation cooling, and neutrino
emission can be more complicated due to strong pion/muon
cooling, so a more careful study is needed to obtain
quantitative constrains on $\eta_p$ in this regime.
On the other hand,
the larger dissipation radius leads to lower neutrino flux and
higher cooling break energy according to equations (12) and (13).
The shift of the first break to higher energies for larger
dissipation radii is due to that those GRBs with $\alpha>1$, whose
neutrino spectral peaks locate at the cooling breaks, contribute dominantly
to the neutrino flux.

\subsection{Uncertainty in the bulk Lorentz factor}
In the previous subsections, we took either the variability or dissipation
radius as a principal parameter, given a Lorentz factor, i.e. $\Gamma=10^{2.5}$.
For those bursts without measured
redshift, we took $L_\gamma=10^{52}{\rm erg\, s^{-1}}$ for the peak
luminosity, as done by ICC. However, it was found recently that, the
bulk Lorentz factor could vary significantly among bursts, and there
is an inherent relation between the Lorentz factor and the isotropic
energy or the peak luminosity (Liang et al. 2011; Ghirlanda et. al.
2011).  As shown by Equations (\ref{nunnupion}) and (\ref{nunnumuon}),
the neutrino flux is very sensitive to the bulk Lorentz factor, so
we can use the inherent relation to obtain more realistic values for
the Lorentz factors and, hence, more reliable estimate of the
neutrino flux.

By identifying the onset time of the forward shock from the optical
afterglow observations, Liang et al. (2011) and Lv et al. (2011)
obtain the bulk Lorentz factors for a sample of GRBs. They further
found a correlation between the bulk Lorentz factor and the
isotropic energy of the burst, given by\footnote{We adopt only the
center value for the relationships presented hereafter.}
\begin{equation}\label{L:G-E}
\Gamma_L=118E_{\rm iso,52}^{0.26}.
\end{equation}
Ghirlanda et. al. (2011) revisit this problem with a large sample
and obtain a relation as
\begin{equation}\label{G:G-E}
\Gamma_G=29.8E_{\rm iso,52}^{0.51}.
\end{equation}
Compared with the benchmark model which assumes $\Gamma=10^{2.5}$
for all bursts, the value of $\Gamma$ obtained from these relations
is lower for bursts with the isotropic energy $E_{\rm
iso}\la(4.4-9.4)\times10^{53}{\rm erg}$.

Ghirlanda et. al. (2011) also obtained  the relation between the
bulk Lorentz factor and peak luminosity, i.e.
\begin{equation}\label{G:L-G}
\Gamma_{G_L}=72.1L_{\gamma,52}^{0.49}.
\end{equation}
The value of $\Gamma$ obtained from this relation is lower than
$10^{2.5}$ for bursts with luminosity $L_p<2.0\times10^{53}{\rm erg
s^{-1}}$.

Yonetoku et al. (2004) and Ghirlanda et al. (2011) found an inherent
relation between the peak energy of photon spectrum and the peak
luminosity. Therefore one can obtain the peak luminosity from the
observed break energy of the photon spectrum
$\epsilon_{\gamma, b}^{\rm ob}$ and the redshift $z$ by adopting the
$\epsilon_{\gamma, b}^{\rm ob}-L_\gamma$ relation, which is
\begin{equation}\label{G:L-e}
L_{\gamma G,52}=7.54\left[\epsilon_{\gamma {\rm b,MeV}}^{\rm ob}(1+z)\right]^{1.75},
\end{equation}
derived by Ghirlanda et. al. (2011).

We use the above inherent relations to calculate both the Lorentz
factor $\Gamma$ and the peak luminosity $L_{\gamma}$, and then
calculate the neutrino flux produced by  the same 215 GRBs, which is
shown in Figure 4. The main differences in the neutrino spectrum
resulted from using different choices of the value of the bulk
Lorentz factor can be summarized as:

(i) The peak energy of the neutrino spectrum shifts to lower energy
for the models adopting the relations in Ghirlanda et al. (2011)
and Lv et al. (2011). This is due to that for the majority
in the 215 GRBs, the values of $\Gamma$ derived with these inherent
relations are lower than the benchmark value $10^{2.5}$, which leads
to a lower peak energy in the neutrino spectrum according to equation
(\ref{enp}). Also, the cutoff energy shifts to lower energies for
these models according to equations (\ref{encpion}) and (\ref{encmuon}).

(ii) The peak flux of the neutrino spectrum increases for the two
models which adopt the inherent relations. This is due to that, a
lower Lorentz factor leads to higher neutrino production efficiency
in the internal shock model. The predicted neutrino flux for both
models adopting $E_{\rm iso}-\Gamma$ relations
in Ghirlanda et al. (2011) and Lv et al. (2011)
exceed the IceCube upper limit, which implies a baryon
ratio $\eta_p\la10$ if $t_{\rm v}^{\rm ob}=0.01$s for long GRBs is correct.

In the left panel of Figure 4, we take the redshift $z=2.15$ for
those long GRBs without measured redshift, the amount of which is
about 84$\%$ of the total amount of GRBs. For the benchmark model,
the neutrino flux would not be affected significantly since it is
independent of the redshift according to equations (\ref{nunnupion})
and (\ref{nunnumuon}). However, the value of  redshift can affect
the flux of the neutrino spectrum for models adopting the inherent
relations.
For a fixed observed fluence of the $\gamma-$ray emission, a smaller
redshift will lead to a smaller peak luminosity or isotropic energy.
As a result, the Lorentz factor, derived from the inherent
relations, will be lower, which leads to a higher neutrino
production efficiency. As shown in the right panel of Figure 4, the
flux for the two models adopting inherent relations increase if we
take $z=1$ for those long GRBs without measured redshifts.

In the above discussions, we have implicitly assumed the baryon
ratio $\eta_p=10$ for GRBs, in accordance with the notation
$f_e=0.1$ in Abbasi et al. (2010). This value comes from the
assumption that the radiation efficiency for GRBs is typically 0.1
and  that most of the dissipated internal energy goes into the
accelerated protons. However, the fraction of energy in protons, and
hence the value of $\eta_p$, are not well-known. The null result of
IceCube observations allows us to put some constraints on this
value. Define $\eta_{p,c}$ as the critic value, above which GRBs
would be detected by the corresponding IceCube configurations. In
Table 1, we list the corresponding value of $\eta_{p,c}$ for the
combined IC40 and IC59 analysis. However, one should keep in mind
that $\eta_{p,c}$ depends on the choice of $t^{\rm ob}_{\rm v}$,
so it should be larger for larger values of $t_{\rm v}^{\rm ob}$.

\section{Diffuse neutrino emission from GRBs }
Recently, IceCube also reported observations of diffuse neutrinos by
the 40-string configuration. The non-detection yields an upper limit
of 8.9$\times 10^{-9}$GeV cm$^{-1}$s$^{-1}$sr$^{-1}$, for the
diffuse neutrino flux assuming an $E^{-2}$ neutrino spectrum (Abbasi
et al. 2011b). The expected diffuse GRB neutrino flux can be
obtained by summing the contributions of all GRBs in the whole
universe. To this aim, we also take into account the number distribution
of GRBs over the luminosity (i.e., the luminosity function) as well as the
number distribution at different redshifts.
Although it is not relevant if the GRB parameters ($\Gamma$ and R) do
not depend on the luminosity, it affects results when one adopts specific relations
such as equations (\ref{L:G-E}) or (\ref{G:G-E}) or (\ref{G:L-e}).

We employ three different luminosity functions and the corresponding
source density evolution functions to describe the distribution of
the GRB number over luminosity and redshift. One luminosity function
is suggested by Liang et al. (2007, hereafter LF-L),
\begin{equation}
\frac{dN}{dL_\gamma}=\rho_0
\Phi_0\left[\left(\frac{L_\gamma}{L_{\gamma \rm
b}}\right)^{\alpha_1} +\left(\frac{L_\gamma}{L_{\gamma \rm
b}}\right)^{\alpha_2}\right]^{-1},
\end{equation}
where $\rho_0=1.2\rm Gpc^{-3}yr^{-1}$ is the local event rate of
GRBs, and $\Phi_0$ is a normalization constant to assure the
integral over the luminosity function being equal to the local event
rate $\rho_0$. This luminosity function breaks at $L_{\gamma \rm
b}=2.25\times 10^{52} \rm erg\,s^{-1}$, with indices $\alpha_1=0.65$ and
$\alpha_2=2.3$ for each segment. The normalized number distribution
of GRBs with redshift used by Liang et al. (2007) in obtaining this
luminosity function is (Porciani \& Madau 2001)
\begin{equation}
S(z)=23 \frac{e^{3.4z}}{e^{3.4z}+22.0}.
\end{equation}

Wanderman \& Piran (2009) also suggested a luminosity function in
the form of a broken power--law (hereafter LF-W)
\begin{equation}
\frac{dN}{dL_\gamma}=\rho_0\Phi_0 \left\{
\begin{array}{ll}
(\frac{L_\gamma}{L_{\rm \gamma b}})^{-\alpha_1} &L<L_{\gamma \rm b},\\
(\frac{L_\gamma}{L_{\rm \gamma b}})^{-\alpha_2} &L\geq L_{\gamma \rm
b},
\end{array}
\right.
\end{equation}
where $\rho_0=1.3$Gpc$^{-3}$yr$^{-1}$, $\alpha_1=1.2$,
$\alpha_2=2.4$ and break luminosity $L_{\gamma \rm b}=10^{52.5}\rm
erg\,s^{-1}$. The corresponding normalized number distribution with
redshift is described by
\begin{equation}
S(z)=\left\{
\begin{array}{ll}
(1+z)^{2.1}  &z<3,\\
(1+z)^{-1.4}  &z\geq 3.
\end{array}
\right.
\end{equation}

Another luminosity function we consider here is suggested by Guetta
\& Piran (2007, hereafter LF-G), which is in the same form as that
of Wanderman \& Piran (2009), but with different parameters, i.e.,
$\rho_0=0.27\rm Gpc^{-3}yr^{-1}$, $\alpha_1=-1.1$, $\alpha_2=-3.0$
and $L_{\gamma\rm b}=2.3\times 10^{51}\rm erg\,s^{-1}$. It implies a
smaller local event rate and fewer GRBs at the high luminosity end.
This luminosity function is obtained based on the assumption that
the rate of GRBs follows the star formation history given by
Rowan-Robinson (1999), i.e.
\begin{equation}
S(z)= \left\{
\begin{array}{ll}
10^{0.75z} &z<1,\\
10^{0.75}  &z\geq 1.
\end{array}
\right.
\end{equation}

Denoting the differential neutrino number generated by a GRB with
luminosity $L_\gamma$ at local redshift $z$ by
$dn_\nu/d\epsilon_{\nu}$, the injection rate of neutrinos per unit
time per comoving volume then can be obtained by
\begin{equation}
\Psi(\epsilon_\nu)=\rho(z)\int\frac{dn_\nu}{d\epsilon_{\nu}}
(L_\gamma,\epsilon_\nu)\frac{dN}{dL_\gamma}(L_\gamma)dL_{\gamma},
\end{equation}
where $\rho(z)\equiv\rho_0S(z)$ is the event rate density in the
rest frame. Considering the cosmological time dilation and the
particle number conservation, a neutrino with energy
$\epsilon_{\nu}$ observed at the Earth must be produced at redshift
$z$ with energy $(1+z)\epsilon_{\nu}$ and
$\Psi_\nu(\epsilon_{\nu}^{\rm ob})d\epsilon_{\nu}^{\rm
ob}=(1+z)\Psi_\nu[(1+z)\epsilon_{\nu}^{\rm ob}]d\epsilon_{\nu}^{\rm
ob}$. The total observed diffuse neutrino flux can then be
integrated over redshift,
\begin{equation}
\frac{dN_{\rm tot}}{d\epsilon_{\nu}^{\rm ob}}= \int_{0}^{z_{\rm
max}} \frac{1}{4\pi}\Psi[(1+z)\epsilon_{\nu}^{\rm ob}]\frac{cdz'}{H(z')}\\
\end{equation}
where
$H(z)=H_0/\sqrt{(1+z)^3\Omega_M+\Omega_\Lambda}$ is the Hubble constant at
redshift z.
Here we set $L_{\gamma,\rm min}=10^{50}\rm erg\,s^{-1}$, $L_{\gamma,\rm
max}=10^{54}\rm erg\,s^{-1}$ and $z_{\rm max}=8$.

The results of diffuse neutrino flux are shown in Figure 5. In this
plot, we show the diffuse muon neutrino spectra for both the case in
which the bulk Lorentz factor of the GRBs follows the inherent
relation suggested by Ghirlanda et al. (2011) (solid lines), as
described in Sec. 3, and the case in which a constant value of
Lorentz factor ($\Gamma=10^{2.5}$) is assumed for all GRBs (dashed
lines). The photon spectrum of GRBs is assumed to be a broken
power-law spectrum described by equation (\ref{ngamma}) with the
$\alpha=1$ and $\beta=2$, and the break energy of photon spectrum is
calculated from the peak luminosity of GRBs via the relationship
shown in equation (\ref{G:L-e}).
One can see from this
plot that using different luminosity functions and associated source
density evolution functions leads to very different flux of diffuse
neutrinos. The expected flux for the LF-L function slightly exceeds
the IC40 upper limit, while for the luminosity functions of LF-W and
LF-G, the predicted flux is undetectable even with one--year full
operation of IceCube. Particularly, the LF-G function results in a
very low flux. This is not only because that the local event rate
for this luminosity function is much lower, but also because that
much more GRBs locate at the low luminosity end, which contribute
lower neutrino flux than more luminous ones. To require the diffuse
neutrino flux not to exceed the IC40 upper limit, we have $\eta_p
\la$8  for the luminosity function of LF-L.
For results in cases where the GRB parameters do not depend on the luminosity,
see Murase $\&$ Nagataki (2006a), Murase et al. (2006), and Gupta $\&$ Zhang (2007).

\section{Conclusions $\&$ Discussions}
The non-detection by the increasingly sensitive detector IceCube has provided
interesting implications for various theoretical predictions of neutrino emission from
GRBs. The IceCube collaboration
reported that the IceCube 40-string and  59-string configurations have
reached the sensitivity below the theoretical expectation, which, if
true, would challenge the view that GRBs could be the sources for
UHECRs. However, as also shown by previous works, we show that the
IceCube collaboration used an overestimated theoretical flux in
comparison with the IceCube instrument limit.  We therefore
revisit the analytic calculation of the neutrino flux,
considering the realistic photon energy distribution in calculating the number
density of fireball photons (instead of using the bolometric luminosity as the
luminosity at the break energy), and using the appropriate normalization for
the proton flux to evaluate the neutrino flux.

Using the modified formulas, we
calculate the expected neutrino flux from the 215 GRBs observed
during the operations of IceCube 40 and 59 strings configurations,
assuming the same benchmark parameters as that used by the IceCube
collaboration.  The flux is about 36$\%$ of the 90$\%$ CL upper
limit, consistent with the non-detection of IceCube for the combined
data analysis of IC40 and IC59.

The benchmark model assumes constant values for the bulk Lorentz
factor, the observed variability time and the peak luminosity for
every burst. Recently, it was suggested that there are correlations
between the bulk Lorentz factor and the isotropic energy, and
between peak luminosity and the break energy of photon spectrum.
Using such inherent relations to derive the Lorentz factor and the
peak luminosity, we re-calculate the neutrino flux and find that
the flux adopting these relations exceed the 90$\%$ CL upper limit
for the assumption of $t_{\rm v}^{\rm ob}=0.01{\rm s}$ for every long
burst. This constrains the baryon ratio to be $\eta_p\la10$, which,
however, could be relaxed if the variability times for most GRBs
could be larger.

We also calculate the cumulative diffuse flux from GRBs using three
different luminosity functions  existing in the literature. For the
luminosity functions of Guetta $\&$ Piran (2007) and Wandermann $\&$
Piran (2009), the  expected flux is below the IceCube upper limit
for both the case that assumes $\Gamma=10^{2.5}$ for every bursts
and the case considering the inherent relation between $\Gamma$ and
the peak luminosity. However, for the luminosity function obtained
in Liang et al. (2007), the expected flux exceeds the IceCube limit
for both cases. The non-detection of diffuse neutrinos then
constrains the baryon ratio to be $\eta_p\la8$ in this case.

GRBs have been proposed to be potential sources for UHECRs, besides
active galactic nuclei (e.g. Biermann $\&$ Strittmatter 1987; Takahara 1990;
Berezinsky et al. 2006) and hypernovae/supernovae with relativistic components
(Wang et al. 2007; Liu \& Wang 2012; Murase et al. 2008). Neutrino detection would
provide evidence for cosmic ray protons in GRBs. On the other hand,
the non-detection by current IceCube can not yet exclude this
connection\footnote{ The argument that the GRB-UHECR connection is
challenged by the IceCube non-detection in Ahlers et al. (2011) is
based on the assumption that   cosmic ray protons are produced by
$\beta$-decay of neutrons from p$\gamma$-interactions that escape
from the magnetic field. }, as the required baryon ratio for GRBs to
be the sources of UHECRs is $\eta_p\simeq 5-10$ (Liu et al. 2011)
for local GRB rate of $R\simeq 1 {\rm Gpc^{-3} yr^{-1}}$ (Liang et
al. 2007; Wanderman \& Piran 2009). Future more sensitive
observations by IceCube or other neutrino telescopes may put more
tight constraints on the baryon ratio and would be able to judge the
GRB-UHECR connection.

We are grateful to Peter Redl, Nathan Whitehorn, Svenja H{\"u}mmer,
Alexander Kusenko, Zhuo Li and Juan Antonio Aguilar for valuable
discussions. This work is supported by the NSFC under grants
10973008, 10873009 and 11033002, the 973 program under grants
2009CB824800  and 2007CB815404, the program of NCET,  the Creative
Research Program for Graduate Students in Jiangsu Province, the Fok
Ying Tung Education Foundation, the Global COE Program. S.N.
acknowledges support from Ministry of Education, Culture, Sports,
Science and Technology (No.23105709), Japan Society for the
Promotion of Science (No. 19104006 and No. 23340069), and the Global
COE Program 'The Next Generation of Physics, Spun from University
and Emergence from MEXT of Japan'. K.M is supported by CCAPP at OSU
and JSPS.

\clearpage
\begin{figure*}
\centering
\plotone{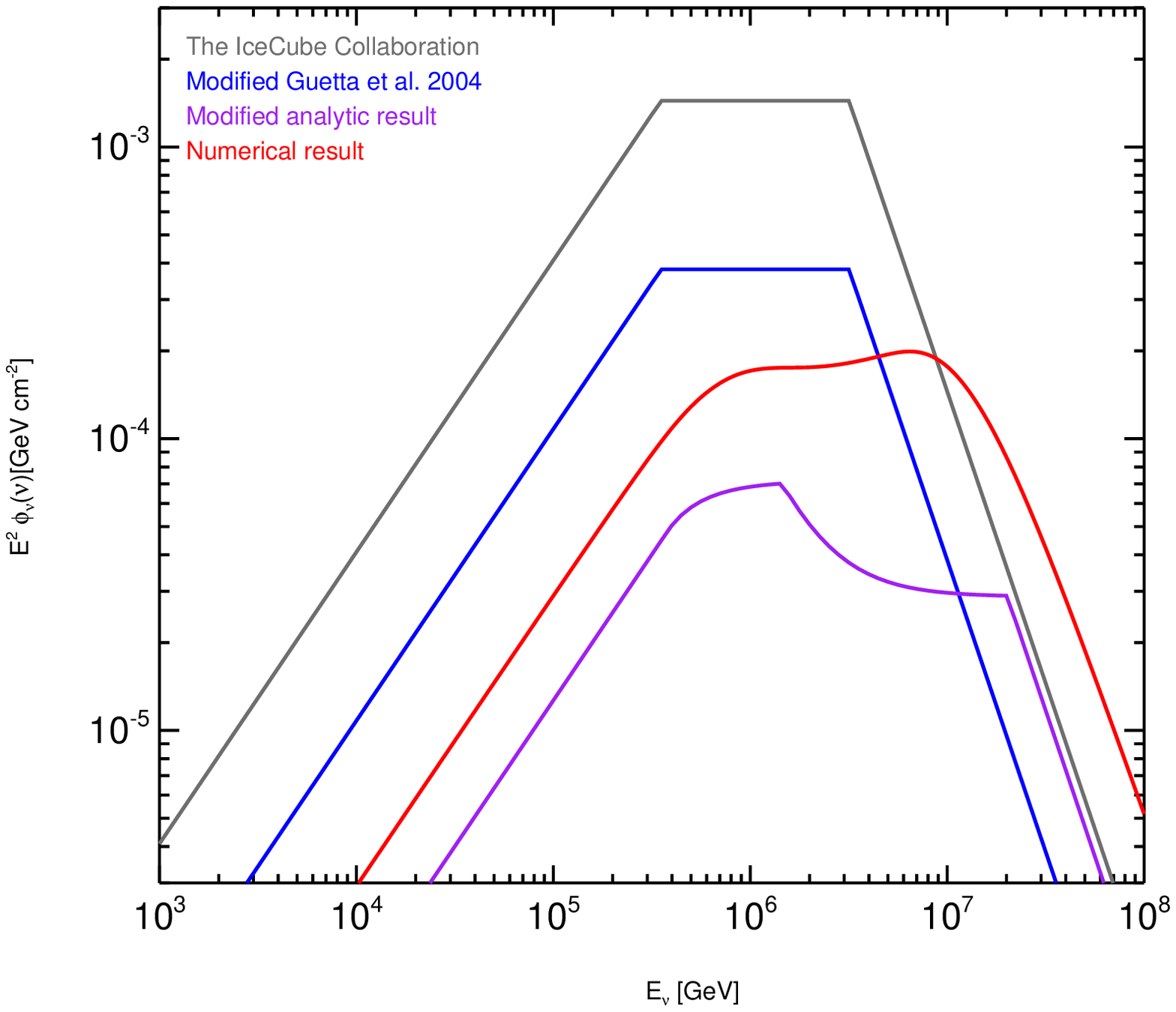} \caption{ The neutrino spectrum for one typical
GRB obtained, respectively, with the method adopted by the IceCube
collaboration (Abbasi et al, 2010, 2011a, the IceCube collaboration, 2011)
(dark gray solid line),
modified Guetta et al. (2004)'s method (blue solid line), our
modified analytical method (purple solid line) and our numerical
method (red solid line). The parameters used in the calculation for
this GRB are: $\alpha=1$, $\beta=2$,  fluence $F_\gamma^{\rm
ob}=10^{-5}{\rm erg\,cm^{-2}}$ (in $10{\rm keV}$ to $1{\rm MeV}$),
$z=2.15$, peak energy $\epsilon_{\gamma,b}^{ob}=200{\rm keV}$, peak
luminosity $L_{\gamma}=10^{52}{\rm erg\,s^{-1}}$,  bulk Lorentz
factor $\Gamma=10^{2.5}$,  the observed variability timescale $t_{\rm
v}^{\rm ob}=0.01{\rm s}$ and the baryon ratio
$\eta_p=10$.
}
\end{figure*}

\begin{figure*}
\plotone{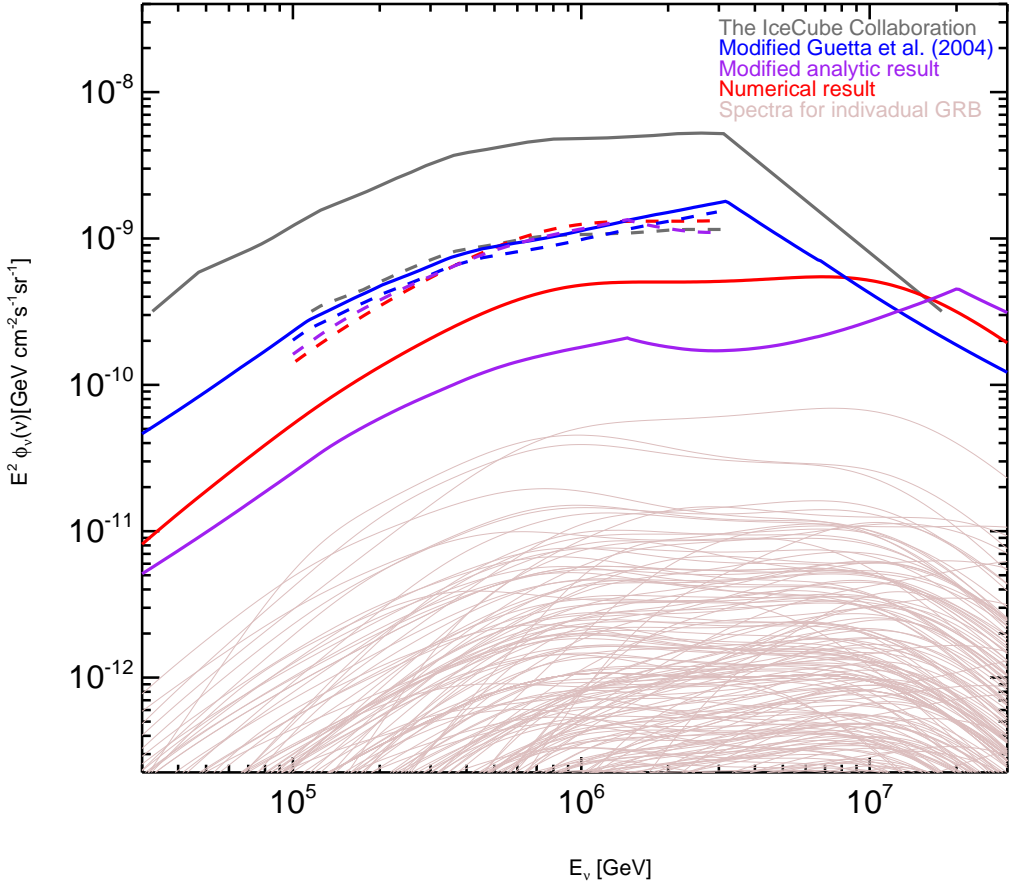} \caption{ Neutrino spectra calculated numerically
by adopting the internal shock radius $R=2\Gamma^2ct_{\rm v}^{\rm
ob}/(1+z)$ for 215 GRBs (light red lines) observed during the
operations in IceCube 40-string and 59-string configurations. We
take the same GRB samples, the same assumptions for GRB parameters
and the same effective area as function of zenith angle as those
used by IceCube collaboration. The red thick solid line represents
the sum of the 215 GRB neutrino spectra and the thick red dashed
line is the corresponding 90$\%$ CL upper limit of IceCube. The
thick dark gray solid line and dashed line are the predicted total
neutrino spectrum and the corresponding 90 $\%$ CL upper limit given
by the IceCube collaboration for the combined data analysis of IC40
and IC59 (The IceCube collaboration 2011). The blue solid and dashed
lines correspond to the expected spectra and 90 $\%$ CL upper limit
obtained by using the modified method in Guetta et al. (2004). The
purple lines represent our modified analytical calculation as a
comparison. For the above calculations, we adopt benchmark
parameters, such as, the peak luminosity $L_\gamma=10^{52}{\rm
erg\,s^{-1}}$, the observed variability timescale $t_{v}^{\rm
ob}=0.01$s for long GRBs,  the Lorentz factor $\Gamma=10^{2.5}$ and
the baryon ratio $\eta_p=10$ for every GRB. }
\end{figure*}

\begin{figure*}
\centering
\plotone{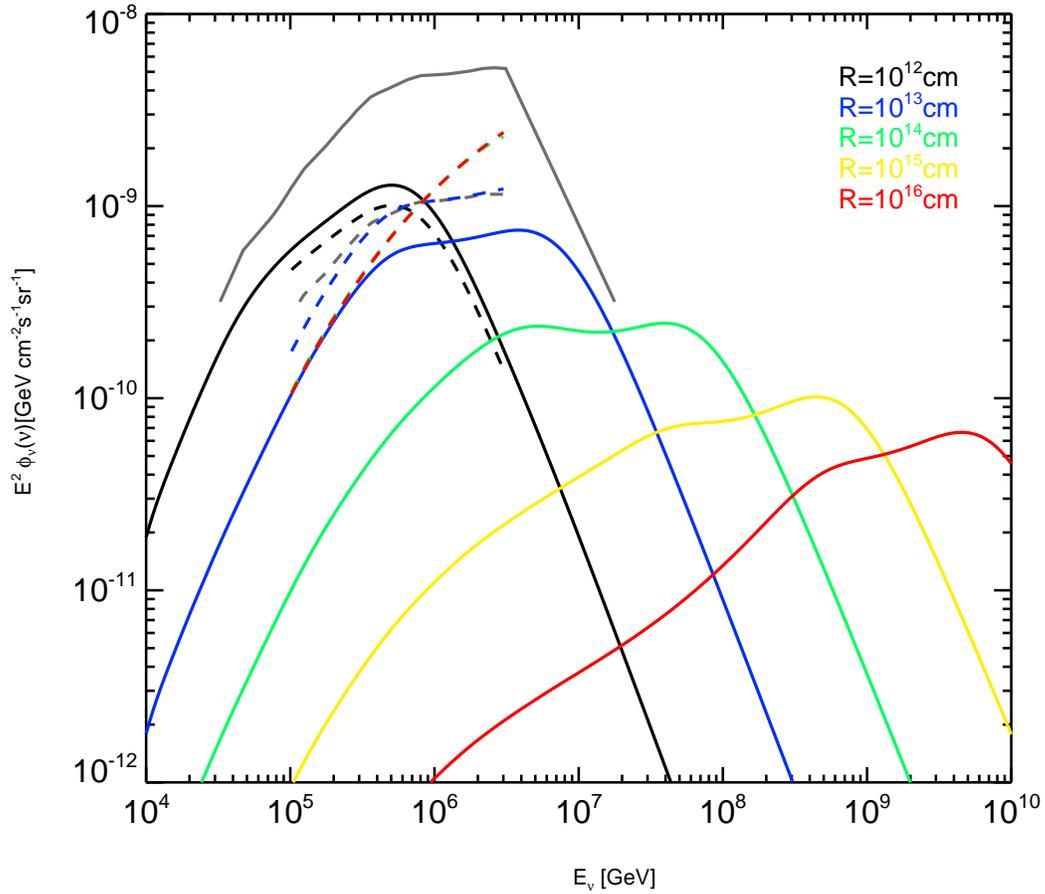} \caption{ The spectra of the total neutrino
emission produced by 215 GRBs assuming the same dissipation radius
for every GRB at $R=10^{12}{\rm cm}$ (the black solid line),
$R=10^{13}{\rm cm}$ (the blue solid line), $R=10^{14}{\rm cm}$ (the
green solid line), $R=10^{15}{\rm cm}$ (the yellow solid
line),$R=10^{16}{\rm cm}$ (the red solid line) respectively. The
corresponding upper limits are shown by the dashed lines. Other
parameters are the same as that used in Figure 2.
Note here, the red, green and yellow dashed lines are overlapped with
each other, because the spectrum shape of the red, green and yellow
solid lines are similar in the energy range of $10^5{\rm GeV}-3\times10^6{\rm GeV}$.}
\end{figure*}

\begin{figure*}
\plotone{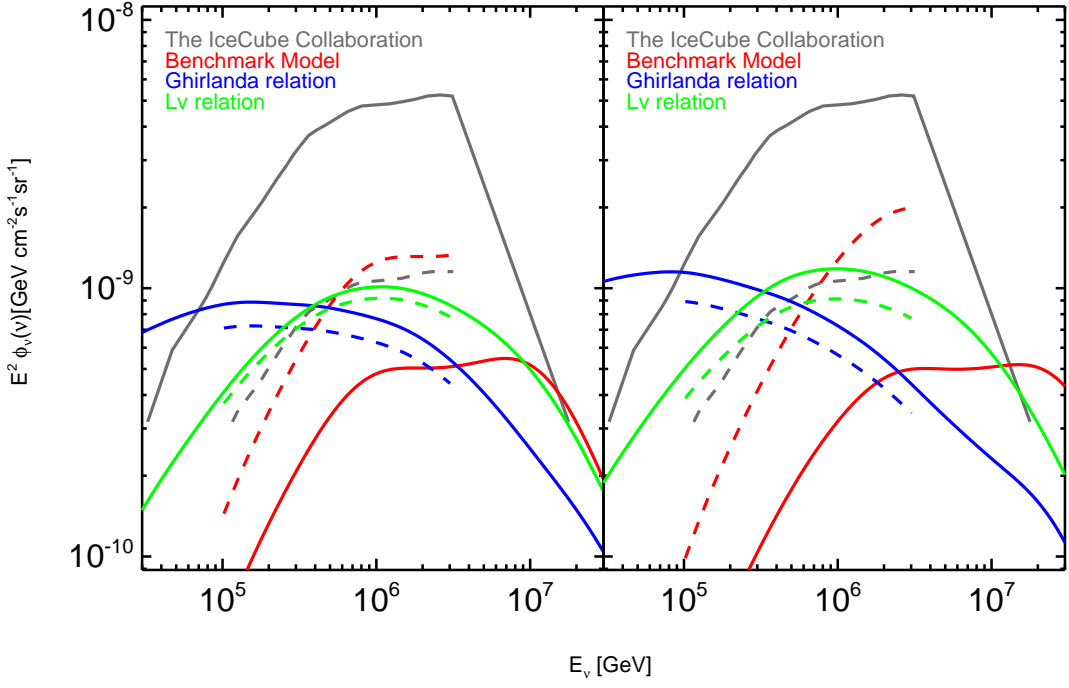} \caption{ The spectra of the total neutrino
emission produced by 215 GRBs assuming different fireball parameters
in the standard internal shock model. The solid red line represents
the spectrum that adopts benchmark parameters as in Figure 2.
The solid blue line represents the spectrum that adopts the
relations of $E_{\rm iso}-\Gamma$ and $\epsilon_{\gamma b}^{\rm
ob}-L_\gamma$ in Ghirlanda et al. (2011). The solid green line
represents the spectrum  that adopts $E_{\rm iso}-\Gamma$ relation in
Lv et al. (2011) and $\epsilon_{\gamma b}^{\rm ob}-L_\gamma$ relation in
Ghirlanda et al. (2011). The dashed lines are the corresponding
upper limit by IC40+IC59. Left panel: $z=2.15$ is assumed for long
GRBs without measured redshifts; Right panel: $z=1$ is assumed
for long GRBs without measured redshifts. }
\end{figure*}

\begin{figure*}
\plotone{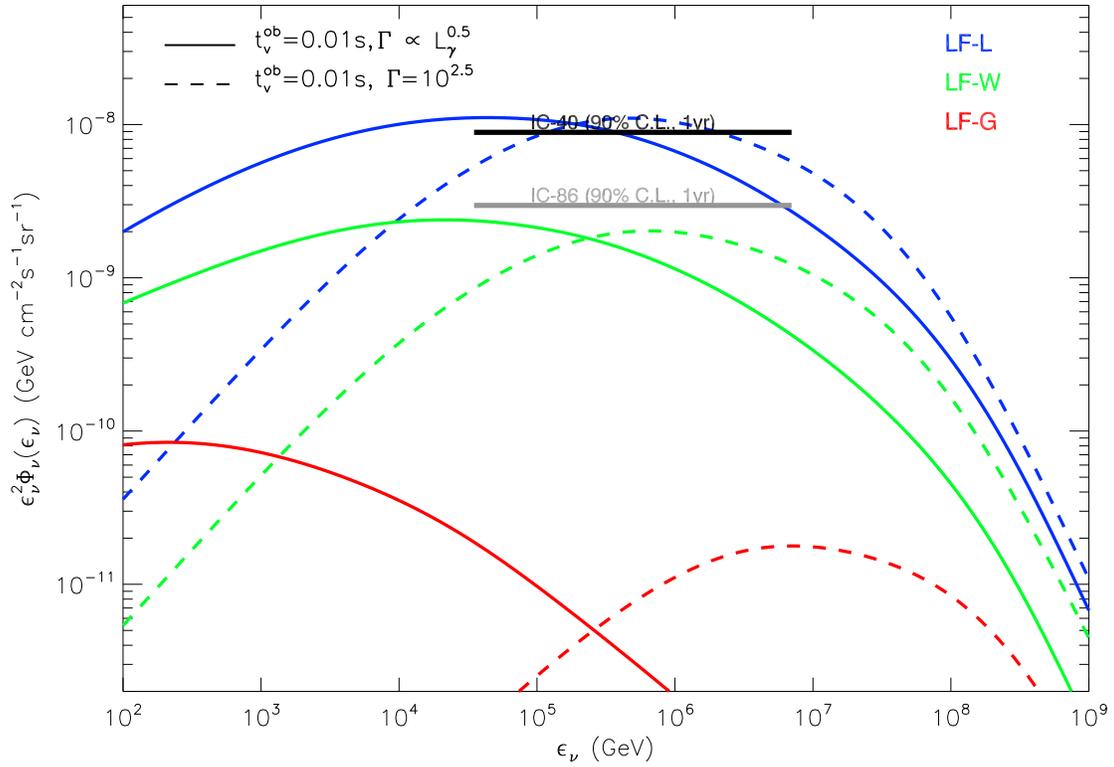} \caption{The expected diffuse muon neutrino flux
from GRBs and the IceCube limits. The blue, green and red lines
represent the fluxes obtained with luminosity functions of Liang et
al. (2007), Wanderman \& Piran (2009) and Guetta \& Piran (2007)
respectively. The solid line and dashed lines correspond to
different assumptions about the Lorentz factor used in the
calculation, but with the same observed variability timescale
$t_{\rm v}^{\rm ob}=0.01$ s for long GRBs and the same baryon loading ratio
$\eta_{p}=10$.
The black thick solid line is the IC40 upper limit on the diffuse
muon neutrino flux given in  Abbasi et al. (2011b), while the dark gray
solid line is the upper limit for one-year observation of the
complete IceCube, extrapolated from the upper limit of IC40
via $A_{\rm eff}^{\rm IC86}\simeq 3A_{\rm eff}^{\rm IC40}$
(Karle 2011, H{\"u}mmer et al. 2011).
 }
\end{figure*}

\begin{table}
\caption{}
  \begin{center}
\label{tab3}
  \begin{tabular}{lllll}

    \hline\hline
               $L_{\gamma}(\rm erg\, s^{-1})$&$\Gamma$&z &$\eta_{p,c}$ \\
                       \hline
           $10^{52}$&$10^{2.5}$ &2.15 &26.0 \\
             &   & 1   & 39.9    \\
           $L_{\gamma\rm G}$& $\Gamma_{\rm G}$& 2.15& 8.16 \\
              &  & 1       &    7.79    \\
       $L_{\gamma\rm G}$& $\Gamma_{\rm L}$ & 2.15&   9.07  \\
             &   & 1   &  7.72\\
    \hline\hline

\end{tabular}
\end{center}
NOTE$-$The critical value of the baryon ratio $\eta_{p.c}$ for the combined
IC40+IC59 analysis obtained
by adopting different assumptions for the bulk Lorentz factor, peak
luminosity and redshift (for long GRBs without measured redshifts).
$L_{\gamma \rm G}$ represents the peak
luminosity obtained by using the  $\epsilon_{\gamma b}^{\rm
ob}-L_\gamma$ relation in Ghirlanda et al. (2011). $\Gamma_{\rm G}$ and
$\Gamma_{\rm L}$ are the Lorentz factors obtained with the relations
of $E_{\rm iso}-\Gamma$ in Lv et al. (2011) and Ghirlanda et al.
(2011) respectively. Here, the observed variability timescale for long GRBs is assumed to be $0.01$ s.
 \end{table}



\end{document}